\documentclass[showpacs,twocolumn,amsmath,amssymb,prb]{revtex4-1}

\usepackage[dvips]{graphicx}
\usepackage{wrapfig,subfigure}
\usepackage{amsmath,amsfonts,amssymb,amsthm}
\usepackage{fancyhdr}
\usepackage{mathrsfs}

\usepackage[normalem]{ulem} 
\usepackage{color} 

\begin{document}

\title{Circuit-QED analogue of a single-atom injection maser: Lasing, trapping states and multistability}

\author{ M. Marthaler,  J. Lepp\"akangas, and J.~H. Cole}

\affiliation{Institut f\"ur Theoretische Festk\"orperphysik
and DFG-Center for Functional Nanostructures (CFN), Karlsruhe Institute of Technology, D-76128 Karlsruhe, Germany}

\begin{abstract}
 We study a superconducting single-electron transistor (SSET) which is coupled to a
 LC-oscillator via the phase difference across one of the Josephson junctions.
 This leads to a strongly anharmonic coupling between the SSET and the oscillator.
 The coupling can oscillate with the number of photons which makes this system very
 similar to the single-atom injection maser. However, the advantage of a design based
 on superconducting circuits is the strong coupling and existence of standard methods
 to measure the radiation field in the oscillator. This makes it possible to study
 many effects that have been predicted for the single-atom injection maser in a
 circuit quantum electrodynamics setup.
 \end{abstract}

\maketitle


{\bf Introduction} The ability to fabricate and control superconducting quantum
 circuits has given birth to the field of circuit quantum electrodynamics (circuit QED).
 Within these circuits, quantized charge, superconducting phase difference and even
 individual microwave photons can be controlled and manipulated, introducing the
 idea of `artificial' atom-photon physics.  Such circuits have been used to reach the
 strong coupling regime\cite{Wallraff2004,UltraStrong}, observe heating and
 cooling\cite{Blencowe2005,Hauss2007,Gracjar2008}, realize a
 three-level-laser\cite{Rodrigues2007,Andre2009,JinStephan,Astafiev2007} and make great strides
 forward in creating highly tunable 'artificial' atoms coupled to an oscillator\cite{CQED}.
 A number of proposals\cite{Marthaler2008,Parametric} and ultimately
 experiments\cite{Astafiev2007,Gracjar2008} have realized the idea of a
 single `artificial-atom' laser (or more precisely maser).  Such work strives towards a
 circuit analogy of the experimental generation of coherent microwave radiation using
 single-atom masers\cite{Filipowicz1986,Lamb1999,Walther1325}.

 Despite much progress, several important physical effects that have been
 painstakingly studied in single-atom injection masers are yet to be realized
 in circuit-QED.  These effects include multistability of the cavity
 field~\cite{Multistability1994}, trapped photon numbers states~\cite{TrappingStates1999}
 and the creation and annihilation of single photons~\cite{PhotonBirthandDeath2007}.  In all these
 experiments, the state of the system is inferred via measurement of the excited
 state occupancy of Rydberg atoms as they drop through a microwave cavity.

  In this paper we will investigate a superconducting single-electron transistor (SSET)
  coupled to a LC-oscillator via the phase difference across the SSET's right-hand
  Josephson junction (JJ).  In doing so, we show that this device allows one to reach
  a regime of strongly nonlinear coupling. It displays multistability
  and trapping states, as well as possessing an operating point which provides protection
  against low-frequency charge noise. As such, this device provides a close analogy to
  the single-atom injection experiments of cavity-QED.  However in contrast to previous
  experiments, a circuit-QED setup allows the direct detection of the photon state of the
  resonator via time resolved measurements of the emitted microwave
  radiation\cite{WallraffCorrelatorMeasurment}.



\begin{figure}
 \begin{center}
 \includegraphics[width = 8 cm]{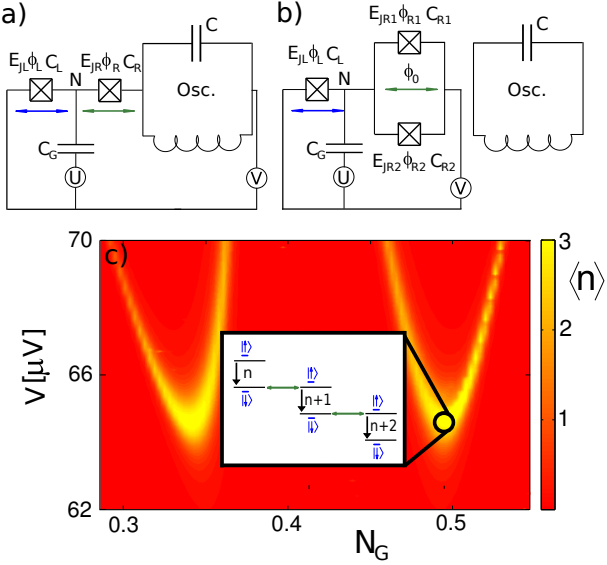}
 \caption{ (a) The superconducting single-electron transistor (SSET) in series with
          the LC-oscillator.
          The SSET consists of two Josephson junctions (crossed boxes)
          with a capacitively connected gate on the small island in between.
           (b) Illustration of the inductive coupling scheme, where
         a SQUID forms an effective right-hand JJ and is thread by the bias flux $\phi_0$
         and the oscillator flux $i{\cal G}(a^{\dagger}-a)$.
         (c) The photon number $ \langle n \rangle$ in the oscillator
         [setup (a)] as a function of the
         gate charge $N_G\propto U$ and the transport voltage $V$.
          Two similar resonance regions  appear corresponding to
          interchanging the roles of the two JJs.
          A symmetry point of particular interest is circled. At this point,
            photons are created in a cascade of energy decay and Cooper-pair
          tunneling through the
          right junction (see text for more detailed explanation).
          For our discussion of the regime of strongly nonlinear coupling,
         we restrict ourself to this symmetry point.
         The parameters of the simulation are
         $E_{JL}=E_{JR}=3\omega_0/10=30$~$\mu$eV, $C_L=C_R=40C_G=C/1250=0.4$~fF
         (${\cal G}=0.08$), $T=50$~mK,
         $R_U=R_V=100$~$\Omega$, and we have included an additional oscillator $Q$-factor \mbox{$Q=2\times 10^3$}.
         }
\label{fig_SSETandschemeexplained}
 \end{center}
\end{figure}

 {\bf The System} In the following we discuss two possibe realizations of our
 desired circuit.  One realization is the series coupling of the
 oscillator with the SSET, Fig.~\ref{fig_SSETandschemeexplained}a. The other
 option is the more conventional inductive coupling scheme\cite{Gracjar2008},
 Fig.~\ref{fig_SSETandschemeexplained}b. In the latter
 it is necessary that a SQUID forms the effective right-hand JJ
 of the SSET. Although incorporating a SQUID will provide additional tunability to the former case as well.
 We divide the coherent Hamiltonian into three parts, describing the artificial
 atom, the photon mode and the interaction between them
  \begin{equation}
   H_{0}=H_{\rm atom}+H_{\rm int}+H_{\rm photon}\, .
   \label{coherent}
  \end{equation}
 In both realizations the artificial atom, that forms the basis of the micro-maser,
 is provided by the charge on the SSET island and Cooper-pair tunneling
 across the left JJ, and is described by the Hamiltonian
 \begin{eqnarray}\label{eq_Effective_Atom_Hamiltonian}
  H_{\rm atom} &=& 4E_C(N-N_G)^2-E_{JL}\cos(\phi_L).
  \label{hamiltonian1}
 \end{eqnarray}
 Here $E_{JL}$ is the corresponding Josephson coupling and
 the charge on the island $2eN$ and the phase difference $\phi_L$ across the left JJ
 satisfy the periodic commutation relation $[N,e^{\pm i\phi_L}]=\pm e^{\pm i\phi_L}$.
 The explicit forms of the gate charge $N_G$ and the charging energy $E_C$ depend on the
 realization and are given below. The gate voltage $U$ is used to bias the system
 such that single-Cooper-pair tunneling is resonant across the left JJ.
 Independent of the realization of the circuit, the quantized mode in a nearby
 stripline or electromagnetic oscillator has the role of the photon cavity and
 is described as a harmonic oscillator ($\hbar=1$), $ H_{\rm photon}=\omega_0 a^{\dagger}a$.

 For the series coupling scheme the interaction Hamiltonian between the atom and the cavity
 has the form
  \begin{eqnarray}
 H_{\rm int}=-E_{JR}\cos[2eVt+\phi_L+{\cal G}(a+a^{\dag})]+H_{\rm cc},
 \end{eqnarray}
 where  $E_{JR}$ is the Josephson coupling across the right-hand JJ and
 we have used the condition that the phase difference across the SSET and the oscillator
 is fixed by the transport voltage $V$, similar to the approach used in
 Ref.~\onlinecite{vandenBrink1991}. Additionally we have a capacitive coupling term,
  \begin{eqnarray}
 H_{\rm cc}=-\frac{2iE_{\rm cc}}{\cal G}N(a^{\dagger}-a).
 \end{eqnarray}
 The important coupling constant is in this case ${\cal G}=(2\epsilon_C/E_L)^{1/4}$,
 where $\epsilon_C=e^2(C_G+C_L+C_R)/2C_{\Sigma}^2$,
 $C_{\Sigma}^2=(C_G+C_L)C_R+C(C_G+C_L+C_R)$ and $E_L$ is the magnetic energy of
 the inductor (the capacitances are defined in Fig.~\ref{fig_SSETandschemeexplained}).
 The charging energy of the SSET island is $E_C=(C+C_R)e^2/2C_{\Sigma}^2$,
 the strength of the capacitive coupling is characterized by
 $E_{\rm cc}=e^2C_R/C_{\Sigma}^2$, and the gate charge has the form
 $N_G=[V(C_R/(C+C_R))(C+C_R(C_G+C)/(C_G+C_L+C_R))-U(C_G-C_R^2
 C_G/((C+C_R)(C_G+C_L+C_R)))]/2e$.
Finally, the frequency of the oscillator is
 $\omega_0=\sqrt{8\epsilon_C E_L}$.


 The  inductive coupling scheme (Fig.~\ref{fig_SSETandschemeexplained}b) results in a similar atom-photon coupling.
 The interaction Hamiltonian is then
  \begin{eqnarray}
 H_{\rm int}
 &=&E_{JR}     \sin\left[2eVt+\phi_L-\frac{i{\cal G}(a^{\dag}-a)}{2}\right]\\
 & & \times    \sin\left[\frac{i{\cal G}(a^{\dag}-a)}{2}\right],\nonumber
 \label{inductivecoupling}
 \end{eqnarray}
 where we assumed equal Josephson energies of the right-hand JJs $E_{JR1}=E_{JR2}=E_{JR}/2$
 and an additional external flux $\phi_0=\pi$ in the SQUID. Here the coupling strength
 ${\cal G}$ depends on the geometry of the system and remains small, ${\cal G}\ll 1$,
 in contrast to the series coupling scheme where ${\cal G} \sim 1$ is possible.
 For the gate charge we obtain $N_G=C_RV-C_GU$ ($C_R=C_{R1}+C_{R2}$) and the charging energy becomes
 $E_C=e^2/2(C_L+C_R+C_G)$.

 Additionally to the coherent time evolution, incoherent
 effects are introduced through voltage fluctuations leading to $V\rightarrow V+V_f$ and
 $U\rightarrow U+U_f$.
 It is this  noise (which is present in any circuit coupled to an electromagnetic
 environment)
 that creates directionality in our lasing cycle,
 since all other processes are coherent and therefore time-reversible.
 The voltage fluctuations are described by the Hamiltonian
 $H_{\rm EE}$ and are mediated on the SSET by the operators
 $N_V$ and $N_U$. The total system is then described by a Hamiltonian of the form~\cite{Leppakangas2008}
 \begin{eqnarray}
  H_T&=&H_{0} + 2eN_VV_f+2eN_UU_f+H_{\rm EE}\, .
 \end{eqnarray}
 We assume that the voltage fluctuations are  characterized by equilibrium
 correlations of the form
 \begin{eqnarray}
  \langle U_f(t)U_f(0) \rangle &=& \frac{R_U}{\pi}
 \int_{-\infty}^{\infty}d\omega\frac{\omega}{1+\left(\frac{\omega}{\omega_c}\right)^2}\frac{e^{-i\omega t}}{1-e^{-\beta\omega}},
 \end{eqnarray}
 where $R_U$ is a typical  line impedance $\sim 100$~$\Omega$ with a cut off
 $\omega_c=1/R_UC_{U}\gg \omega_0$ and that $U_f$ and $V_f$ are uncorrelated.
 For the series LC oscillator one obtains
 \begin{eqnarray}
N_U&=&-\frac{i}{2}\frac{C_GC_R}{C_{\Sigma}^2}\frac{a^{\dagger}-a}{{\cal G}}\\
&+&\left[\frac{C_G(C+C_R)}{C_{\Sigma}^2} -\frac{C_R^2C_G}{C_{\Sigma}^2(C_G+C_L+C_R)} \right]N\nonumber .
 \end{eqnarray}
 Similar relations apply for transport voltage fluctuations $V_f$ and the interaction operator $N_V$.
 As one can see for series coupling we create additional noise in the oscillator as well.
 We will always include this effect by using an oscillator decay rate which will be
 larger than the decay induced only by the voltage fluctuations to allow for additional
 internal or external loss.

  For the remainder of the paper, we focus on the series coupling scheme
 (Fig.~\ref{fig_SSETandschemeexplained}a) as
 this allows a stronger nonlinear coupling between qubit and oscillator, but our
  results apply to both schemes.
  In Fig.~\ref{fig_SSETandschemeexplained}c we plot the average photon number as a
  function of the gate charge $N_G$ and transport voltage $V$ for our device,
  obtained using a full density-matrix simulation of the SSET and oscillator.
  We diagonalize the Hamiltonian~(\ref{coherent})
  using the Floquet expansion of the eigenstates,
 \begin{equation}
  |\psi\rangle =\sum_{n,n',N}c_{n n' N} e^{2i n'eV t}|n\rangle |N\rangle\, ,
 \end{equation}
 where $|n\rangle$ are the photon number states of the oscillator and $|N\rangle$
 are the charge states of the island.
  We then treat the voltage fluctuations perturbatively by expanding the time evolution
  of the resulting density matrix in orders of the coupling to the reservoirs.
  After tracing out the reservoir degrees of freedom we arrive at the
  Bloch-Redfield equation for the reduced density matrix $\rho$ of the  quantum system.
  A detailed discussion of this formalism for the
  SSET and the methods used to solve the
 master equation can be found in Ref.~\onlinecite{Leppakangas2008}.


 {\bf Cascade Resonance} Several resonances appear
  at different regions in the $V-N_G$ plane shown in Fig.~\ref{fig_SSETandschemeexplained}c.
  These points correspond to a resonance condition between levels of the SSET and the oscillator.
  We focus our further
  discussion on the symmetry point at $N_G=1/2$, where the charge states $|N=0\rangle$ and $|N=1\rangle$
  are degenerate.
  At this symmetry point we are able to qualitatively describe features of the system by a simplified Master equation.
  We will now discuss the relevant terms in this 
  equation and explain how a lasing cycle can be achieved.

 The eigenstates of the effective atom  are given by the qubit states
 \begin{eqnarray}\label{eq_Qubit_States}
  |\uparrow\rangle &=&\frac{1}{\sqrt{2}}(|N=0\rangle-|N=1\rangle)\, ,\\
   |\downarrow\rangle &=&\frac{1}{\sqrt{2}}(|N=0\rangle+|N=1\rangle)\, ,\nonumber
 \end{eqnarray}
 with the energy splitting $\Delta E= E_{JL}$. Close to the symmetry point
 the energy changes with $(N_G-1/2)^2$ and is therefore protected to the first order from low-frequency charge noise~\cite{Makhlin2001}.
 In the basis of the qubit states
 we can write the total coherent Hamiltonian as
 \begin{eqnarray}\label{eq:Ham}
  H_0 &=& \frac{1}{2}\! \Delta E\sigma_z +\omega_0 a^{\dag}a \\
  && +\frac{E_{JR}}{4}\!\left[(\sigma_z + i\sigma_y)e^{i[2eVt+{\cal G}(a^{\dag}+a)]}
                                                  + h.c. 
                                             \right]\, ,\nonumber
 \end{eqnarray}
 where we neglected the nonresonant capacitive coupling $H_{\rm cc}$
 and $\sigma_i$ are the Pauli matrices acting on the eigenstates (\ref{eq_Qubit_States}).
 Now we perform a transformation $U=\exp\left[-i 2 eV a^{\dag} a\,t\right]$
  and, after a rotating-wave approximation, this yields
 \begin{equation}\label{eq_Rotating_Wave_Hamiltonian}
  H_{U} = \frac{1}{2} \Delta E\sigma_z-\omega_{\rm eff} a^{\dag}a+\frac{i\,E_{JR}}{2 }\left[\sigma_+ s_+ -\sigma_- s_-\right]\, ,
 \end{equation}
 with an effective oscillator frequency $\omega_{\rm eff}=2eV-\omega_0$ and the operators
 \begin{equation}\label{eq:s}
  s_+=\sum_n \langle n+1|\sin\left[{\cal G} (a^{\dag}+a) \right]|n\rangle |n+1\rangle\langle n|\, ,
 \end{equation}
  and $s_+=s_-^{\dag}$. At the symmetry point that is circled in
  Fig.~\ref{fig_SSETandschemeexplained}c
  we have $2eV>\omega_0$. This means
  that in our rotating frame each photon that is created
  effectively decreases the energy of the system. 

 The resonance condition is given by $\omega_0=2eV-\Delta E$ ($\Delta E=\omega_{\rm eff}$). In this case there is
  a coherent transition between $|\downarrow\rangle |n\rangle$  and   $|\uparrow\rangle |n+1\rangle$. To create an overall increase in the photon number
  we now need an additional incoherent pumping process which is given by the fluctuations of the electromagnetic environment.
   In the basis of the qubit states (\ref{eq_Qubit_States}) the charge operator is
  transverse to the eigenstates,  $N\propto\sigma_x$. This means the charge noise  allows a decay from
  the qubit state $|\uparrow\rangle$ to the qubit ground state $|\downarrow\rangle$. This in turn creates a higher population in the state
  $|\downarrow\rangle$.
  Through this mechanism photons are created in a cascade of energy decay and absorption of voltage quanta.
  As shown in the schematic in Fig.~\ref{fig_SSETandschemeexplained}c, each time a Cooper-pair tunnels through the right junction
  it creates (or absorbs) a photon. The energy
   to keep the cycle going is provided by the voltage source. In the linear (small ${\cal G}$) limit, this scheme is similar to Rabi-Sideband lasing,
  that is well understood in conventional quantum optics \cite{Rautian1961,Wu1977} and has been discussed and demonstrated in circuit QED
  \cite{Hauss2007,Gracjar2008}.

\begin{figure}[t]
 \begin{center}
 \includegraphics[width = 8 cm]{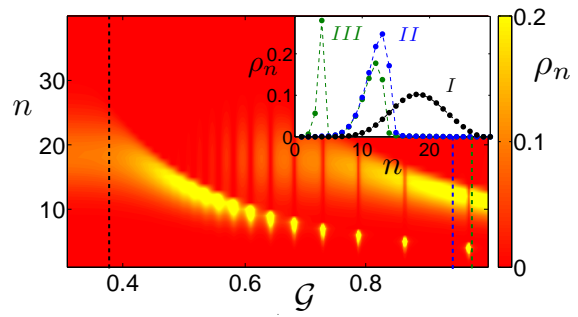}
 \caption{The photon distribution $\rho_n$ as a function of the
           coupling strength $\cal {G}$, where we keep the linear coupling strength
            $E_{JR}{\cal G} $ constant. We see three major types of behavior, each of which we show in the inset.
           A Poisson distribution for small coupling (black, $I$) and for increasing ${\cal G}$ a squeezed distribution (blue, $II$)
            and multistability (green line, $III$). We used the parameters: $\kappa/\gamma=0.027$ and ${\cal G}E_{JR}/\gamma=6.7$.
             }\label{fig_SolutionofMasterEquation}
 \end{center}
\end{figure}

  At the symmetry point we can write the Master equation in a simplified  Lindblad form,
 \begin{equation}\label{eq_MasterEquation}
  \dot{\rho}=-i[H_U,\rho]+{\cal L}_{\rm ch}\rho+{\cal L}_{\rm diss}\rho\, ,
 \end{equation}
 with dissipation in the oscillator
 \begin{eqnarray}
 {\cal L}_{\rm diss}\rho
 &=&
 \frac{\kappa}{2}
 \left(2a\rho a^{\dag}-[a^{\dag}a,\rho]_+\right)\, ,
 \end{eqnarray}
 and qubit decay
 \begin{eqnarray}
  {\cal L}_{\rm ch}\rho=\frac{\gamma}{2}\left(2\sigma_-\rho\sigma_+-[\sigma_+\sigma_-,\rho]_+ \right)\, ,
 \end{eqnarray}
 where $[,]_+$ is the anticommutator, and the decay rate is given by
 \begin{eqnarray}
  \gamma&=&|\langle\downarrow |2eN_U|\uparrow\rangle|^2\int_{-\infty}^{\infty} dt \langle U_f(t)U_f(0) \rangle e^{i \Delta E\, t}\\
        & &+ |\langle\downarrow |2eN_V|\uparrow\rangle|^2\int_{-\infty}^{\infty} dt \langle V_f(t)V_f(0) \rangle e^{i \Delta E\, t}\nonumber .
 \end{eqnarray}
 The overall form of our master equation is now significantly simplified and similar to
 the master equation of a single-atom injection maser. In the rotating-wave approximation
 used to derive the Hamiltonian (\ref{eq_Rotating_Wave_Hamiltonian}) we neglected many matrix elements
 connecting different photon number states. However, comparison to our full solution shows that
 all qualitative features of the system are well described by Eq.~(\ref{eq_MasterEquation}).

{\bf Anharmonic Coupling}
The key difference between this circuit and existing circuit-QED setups is
 the anharmonic coupling term, $\sin\left[{\cal G} (a^{\dag}+a) \right]$, see Eqs.~(\ref{eq:Ham}-\ref{eq:s}).
  To understand the effects of this term, we consider three different regimes.
 Expanding the anharmonic term for weak coupling $\mathcal{G} \ll 1$, we obtain the usual constant coupling term
\begin{equation}
\sin\left[{\cal G} (a^{\dag}+a) \right]= {\cal G} (a^{\dag}+a) + \mathcal{O}\left(\left[{\cal G}(a+a^{\dag})\right]^3\right)\, ,
\end{equation}
which gives in the lowest order $s_{+}={\cal G}a^{\dag}$, resulting in the
 usual single-qubit lasing results~\cite{Hauss2007,Rodrigues2007,Andre2009}. In this case we find
 the photon number at resonance, $\omega_{\rm eff}=\Delta E=E_{JL}$, to be \cite{Scully}
\begin{equation}
  \langle n \rangle_0 = \frac{\gamma}{2\kappa}-\frac{\gamma^2}{2 (E_{JR} {\cal G})^2}\, ,
\end{equation}
 for $\langle n \rangle_0\gg 1$.
 This expression for the photon number in the linear
 coupling limit will prove to be useful in understanding the other regimes that
 we can reach using this device.

 A main advantage of this circuit is that we are
 not limited to the linear coupling regime.  As the strength
 of the effective coupling term is roughly sinusoidal with $\mathcal{G}$
 and photon number, we see a number of new effects.  Ironically,
 when operating in this mode, our micromaser behaves in a very similar
 fashion to a single-atom injection maser~\cite{Filipowicz1986,Multistability1994,TrappingStates1999},
 where our oscillatory coupling provides a direct analog to the spatially
 dependent atom-field coupling in injection masers.  In Fig.~\ref{fig_SolutionofMasterEquation}
 one can see the probability distribution of the photon number states,
 \begin{equation}
  \rho_n=\langle n| \, \mathrm{Tr}_{\uparrow/\downarrow}[ \rho ] \, |n\rangle\, ,
 \end{equation}
 in the stationary limit $\dot{\rho}=0$, as function of the coupling strength $\cal{G}$. As discussed above,
 for small coupling (marked by $I$) we simply get the Poisson distribution we would expect for a laser \cite{Scully}.

 For slightly larger coupling (or photon number) we obtained
 a squeezed number distribution within the oscillator.
 This regime ($II$) is reached when $\pi/2 \lesssim \mathcal{G} \sqrt{\langle n \rangle_0} \lesssim \pi$.
  If this condition is fulfilled, as the photon number approaches the upper
 limit ($\mathcal{G} \sqrt{\langle n \rangle_0} \rightarrow \pi$), the matrix
 element $s_+$ is effectively `cut off' resulting in the usual
 squeezed state physics~\cite{Parametric,Marthaler2008}.  This can be
 seen in the asymmetric character of the photon number
 distribution (the blue curve) in the inset to Fig.~\ref{fig_SolutionofMasterEquation}.

Increasing the coupling
further ($\mathcal{G} \sqrt{\langle n \rangle_0} > \pi$), we
reach a regime of multistability ($III$).  In this regime,
 the system can occupy either the original
 squeezed state, restricted by the first
zero crossing of $\sin\left[{\cal G} (a^{\dag}+a) \right]$,
 or can occupy a new squeezed state associated
with the second or subsequent zeros in the sine function.
 Which of these states are occupied depends on which matrix
 elements are small, resulting in `hot spots'
where the system (as a function of ${\cal{G}}$)
 suddenly jumps from one squeezed state to another,
 as can be seen in the bright regions of
Fig.~\ref{fig_SolutionofMasterEquation}.
  Near the crossover between these states,
the system displays bistability (and at higher $\cal{G}$ multistability)
where the oscillator is in a statistical mixture of
two squeezed states with distinct photon number
distributions (the green curve in the inset of Fig.~\ref{fig_SolutionofMasterEquation}).

\begin{figure}[t!]
 \begin{center}
 \includegraphics[width = 8 cm]{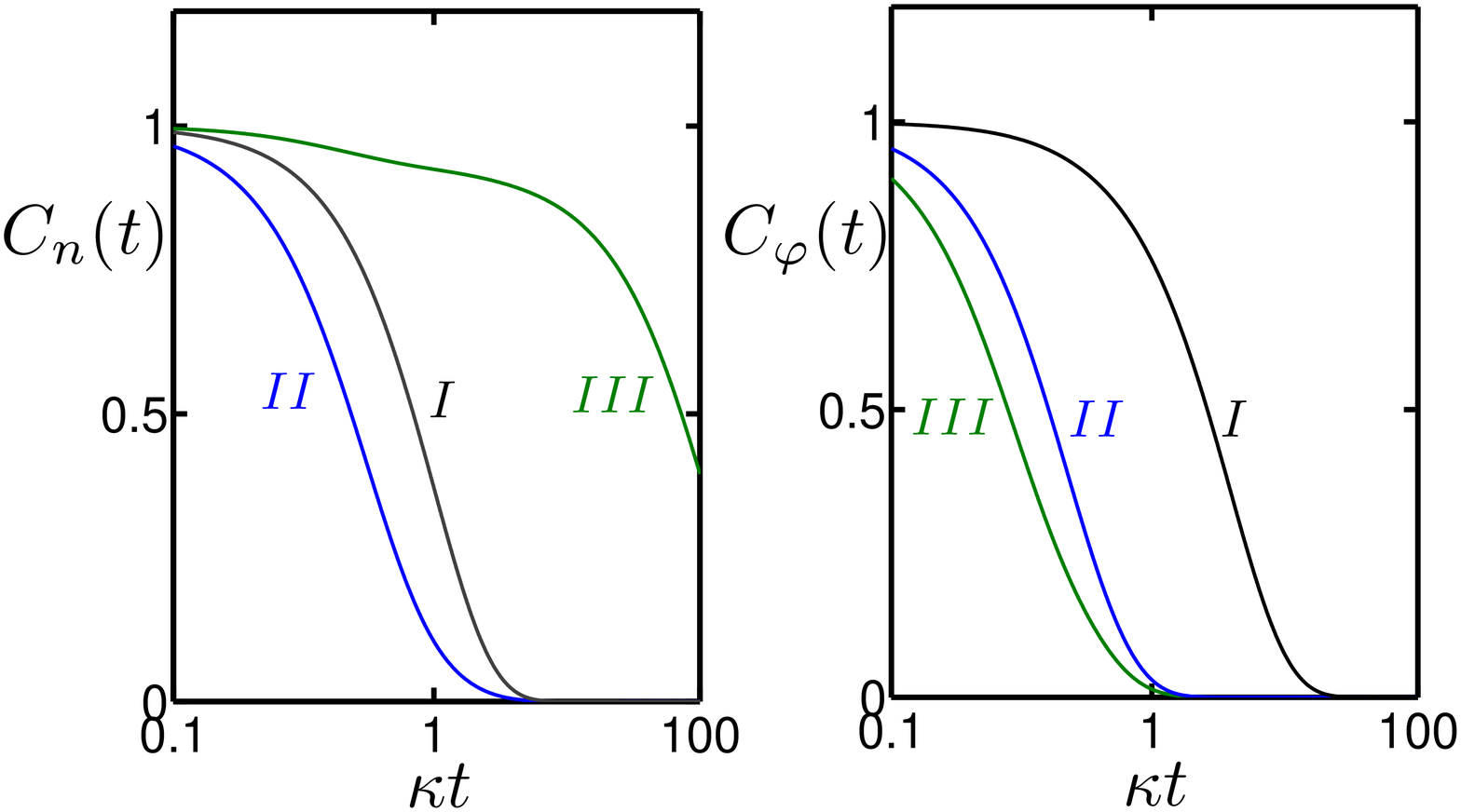}
 \caption{The amplitude $C_{n}$ and the phase correlation function $C_{\varphi}$ as a function of time,
 calculated from a numerical solution of the time evolution of the master equation (\ref{eq_MasterEquation}).
 The different lines correspond to the different operating regimes shown in Fig. \ref{fig_SolutionofMasterEquation}.
 (black line $I$) ${\cal G}=0.37$, (blue line $II$) ${\cal G}=93$, (green line $III$) ${\cal G}=0.968$.
   We used the parameters: $\kappa/\gamma=0.027$ and ${\cal G}E_{JR}/\gamma=6.7$.
}\label{fig_Correlators}
 \end{center}
\end{figure}

{\bf Measurement} A great advantage of the circuit-QED realization of a micromaser is that it
 is possible to make a time resolved measurement of the  microwave radiation emitted
 from the LC-oscillator.
 Measuring the phase and the amplitude fluctuations makes it possible to distinguish between the
 different states of radiation in the cavity. Amplitude fluctuations are described by the amplitude correlation function,
 \begin{equation}
  C_n(t)=\frac{\langle a^{\dag}(t)a(t)a^{\dag}(0)a(0)\rangle-\langle a^{\dag}a\rangle^2}{\langle (a^{\dag}a)^2 \rangle-\langle a^{\dag} a\rangle ^2}\, ,
 \end{equation}
 while phase fluctuations are given by the phase correlation function
  \begin{equation}
  C_{\varphi}(t)=\frac{\langle a^{\dag}(t)a(0)\rangle}{\langle a^{\dag}a \rangle}\, .
 \end{equation}
 Both correlators have been normalized such that \mbox{$C_i(0)=1$}  and they
 approach zero for long times, $C_i(t\rightarrow \infty)\rightarrow 0$.
 With this normalization, each correlator is fully characterized by a single decay rate,
 $C_i(t)\propto e^{-\kappa_i t}$.
  In Fig.~\ref{fig_Correlators} we
 show the amplitude and the phase correlation function for the LC oscillator in the various
 operating regimes. One should note that the time-axis is logarithmic since differing regimes
 produce decay rates that differ by several orders of magnitude.

 The two operating regimes $II$ and $III$ can be clearly distinguished in the amplitude correlation function.
 For a squeezed distribution ($II$), the amplitude is very similar at all times and therefore the amplitude
 decays very fast to its long time limit. If the photon distribution is multistabile the situation is very different.
 In this case the field is fluctuating between two favored photon number states that are only connected by a small matrix element.
 Therefore it takes a long time until the amplitude correlator decays to its stationary value. The decay rates of the
 amplitude correlation function are therefore ordered such that $\kappa_n^{III}>\kappa_n^{I}>\kappa_n^{II} $.
 The decay rate of the coherent state lies between the two other decay rates as it is given approximately by the oscillator decay rate
 $\kappa_n^I \approx \kappa $.

 To find a clear signature distinguishing the coherent state it is necessary to turn to the phase correlation function.
 For the squeezed and the multistable state the phase is not well defined and the correlator decays quickly, $\kappa_{\varphi}^{II,III}\approx \kappa$.
 In contrast, the coherent state has a very long phase correlation time. This is the standard result for a laser \cite{Scully},
 since the phase correlation decay rate for a laser is inversely proportional to the average photon number, $\kappa_{\varphi} \propto 1/\langle n \rangle$,

{\bf Conclusion} We have presented a circuit design for creating non-trivial
microwave photon distributions using a SSET strongly coupled
to a strip-line or LC resonator.  Using a combination of Cooper-pair
tunneling and coupling to voltage fluctuation noise, a lasing cycle can be
established.
As this can be achieved while operating at a charge degeneracy point,
the system is also protected against low-frequency charge noise.

Realizing the strongly nonlinear coupling regime requires an oscillator with
a large ratio of charging energy to inductive energy, ${\cal
G}=(2\epsilon_C/E_L)^{1/4}$. Such an oscillator has been demonstrated in
Ref.~\onlinecite{PaperWithLargeInductivity_Beltran2007}.  A superconducting
quantum interference device array was used as an effective resonator,
which has the additional advantage of allowing for a tunable
$\cal G$, such that all regimes
discussed in this paper can be accessed in a single experiment. Since the
overall photon number
can remain small the anharmonicity of the SQUID array should not play 
significant role.
 Another possible realization is to use a tunable right-hand (or left-hand) Josephson junction. This would
allow to change the photon number of the linear limit
 $\langle n \rangle_0$. An increase of the photon number would allow the
measurement of all three regimes as well.


\end{document}